\documentclass[twocolumn,prb, showpacs]{revtex4-1}
\usepackage{graphicx}
\usepackage{dcolumn}
\usepackage{bm}

\begin{document}

\title{Electronic phase diagram of  Li$_{x}$CoO$_2$ revisited with potentiostatically de-intercalated single crystals}

\author{T. Y. Ou-Yang$^{1,2}$}
\author{F. -T. Huang$^1$}
\author{G. J. Shu$^{1}$}
\author{W. L. Lee$^3$}
\author{M. -W. Chu$^1$}
\author{H. L. Liu$^{2}$}
\author{F. C. Chou$^{1,4}$}
\email{fcchou@ntu.edu.tw}

\affiliation{
$^1$Center for Condensed Matter Sciences, National Taiwan University, Taipei 10617, Taiwan}
\affiliation{
$^2$Department of Physics, National Taiwan Normal University, Taipei 11677, Taiwan}
\affiliation{
$^3$Institute of Physics, Academia Sinica, Taipei 11529, Taiwan}
\affiliation{
$^4$National Synchrotron Radiation Research Center, Hsinchu 30076, Taiwan}

\date{\today}

\begin{abstract}

Electronic phase diagram of Li$_x$CoO$_2$ has been re-examined using potentiostatically de-intercalated single crystal samples.  Stable phases of x $\sim$ 0.87, 0.72, 0.53, 0.50, 0.43, and 0.33 were found and isolated for physical property studies.  A-type and chain-type antiferromagnetic orderings have been suggested from magnetic susceptibility measurement results in x $\sim$ 0.87 and 0.50 below $\sim$ 10K and 200K, respectively, similar to those found in Na$_x$CoO$_2$ system.  There is no Li vacancy superlattice ordering observed at room temperature for the electronically stable phase Li$_{0.72}$CoO$_2$ as revealed by synchrotron X-ray Laue diffraction.  The peculiar magnetic anomaly near $\sim$175K as often found in powder samples of x$\sim$0.46-0.78 cannot be isolated through this single crystal potentiostatic method, which supports the previously proposed explanation to be surface stabilized phase of significant thermal hysteresis and aging character.  

\end{abstract}


\maketitle



\section{\label{sec:level1}Introduction\protect\\ }

\begin{figure}
\begin{center}
\includegraphics[width=3.5in]{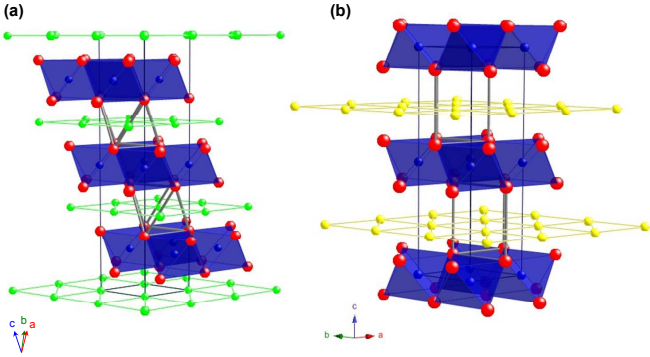}
\end{center}
\caption{\label{fig:fig-structure}(color online) Crystal structures of (a) O3-Li$_x$CoO$_2$ described by rhombohedral R$\bar{3}$m or trigonal layer symmetry and (b) P2-Na$_x$CoO$_2$ described by P6$_3$/mmc symmetry are compared.  Note that Li$_x$CoO$_2$ has three glide layers of LiO$_6$ octahedral per unit as drawn in hexagonal form, and there are two reversed layers of NaO$_6$ trigonal prismatic per unit. }
\end{figure}

Both layered P2-Na$_x$CoO$_2$ (also called $\gamma$-phase) and O3-Li$_x$CoO$_2$ (also called $\alpha$-phase) compounds have been investigated intensely as a rechargeable battery electrode material, and the latter, in particular,  has been commercialized, based on its  proper application voltage, high energy density, and high rate capacity.\cite{Goodenough2010}  In general, both systems are composed of alternating layers of CoO$_2$ and alkali metal with 2D hexagonal symmetry.  As illustrated in Fig.~\ref{fig:fig-structure}, the major difference lies in the oxygen coordination near the cations, i.e., P2-Na$_x$CoO$_2$ can be described by P6$_3$/mmc space group with two opposite prismatic NaO$_6$ layers (P2) sandwiched between Co-layers per unit, while Li$_x$CoO$_2$ shows rhombohedral R$\overline{3}$m or trigonal layer symmetry with three gliding layers of LiO$_6$ octahedra (O3) sandwiched in between Co-layers per unit when drawn in hexagonal form.\cite{Amatucci1996}  The surprising finding of superconductivity in Na$_x$CoO$_2$$\cdot$yH$_2$O (x $\sim$ 1/3 and y $\sim$ 4/3) has generated intense investigation on the rich electronic phase digram of Na$_x$CoO$_2$, including antiferromagnetic ordering, metal-insulator transition, and superconductivity.\cite{Takada2003, Foo2004}  On the other hand, detailed Li$_x$CoO$_2$ electronic phase diagram remains to be illusive while being hindered by the availability of sizable high quality single crystal samples with well-controlled homogeneous Li content.

Na vacancy cluster ordering in layered Na$_x$CoO$_2$ has been shown to be a fascinating phenomenon since the various superstructures were observed through neutron and synchrotron X-ray diffraction techniques.\cite{Roger2007, Chou2008}  The strong correlation between vacancy cluster ordering pattern and the distinct physical properties at specific x values has been explored in detailed since.\cite{Shu2010, Shu2009}  However, although similar characteristic surface potential V vs. x behaviors have been found in the nearly isostructural Li$_x$CoO$_2$ through repeated electrochemical galvanostatic charge/discharge scans, both in-situ and ex-situ, the expected similar Li vacancy orderings from the nearly isostructural Li$_x$CoO$_2$ have been implied yet never observed with solid evidence.\cite{Amatucci1996, Hertz2008}  While the existence of electronically stable phases revealed by the characteristic surface potential do not mean that the preferred specific charge levels would lead to specific Li-vacancy ordering, it is still possible that similar Li vacancy ordering does exist, except that it is smeared at room temperature as a result of thermal fluctuation.  In fact, the first-principles calculation predicted that most stable phases of Li vacancy orderings occur only below x=1/2.\cite{VanderVen1998}  In this report, we have confirmed that no Li vacancy ordering exists for x $\sim$ 0.72 at room temperature, based on synchrotron X-ray Laue diffraction.

While studies on the structure and electrochemical properties of Li$_x$CoO$_2$ using polycrystalline sample for the purpose of battery application, physical property exploration using single crystal samples is rare due to its scarce availability.  Most of the electronic phase diagrams mapped so far were based on powder samples.\cite{Mukai2007, Sugiyama2005, Hertz2008, Motohashi2009} On the other hand, the very few reports based on single crystal samples used either chemical de-intercalation or electrochemical galvanostatic de-intercalation methods,\cite{Miyoshi2009}  which we will argue in the following as having potential multi-phase issue, not to mention the conclusions drawn from single crystal Li$_x$CoO$_2$ samples prepared through  ion-exchange route by P2-Na$_x$CoO$_2$.\cite{Miyoshi2010}  In fact, it is highly likely that the inclusion of O2-Li$_x$CoO$_2$, instead of the expected O3-Li$_x$CoO$_2$, may have occurred through ion-exchanged P2-Na$_x$CoO$_2$ original.\cite{Carlier2001} 

Detailed studies based on single crystal sample through the electrochemical potentiostatic de-intercalation route have not been reported so far.   The particle size difference and packing density could affect the charging efficiency and surface potential readings for the study based on polycrystalline samples.  The common practice of galvanostatic charging based on C-rate in the battery study, i.e., charging rate set on a fraction of the capacity, becomes impractical when there are particularly stable phases existing, instead of a complete solid solution for the ion intercalation.  In order to study the electronic phase diagram of Li$_x$CoO$_2$ accurately, we report phase studies based on single crystal samples that have been prepared electrochemically using the potentiostatic route, similar to the techniques that have been applied successfully to the P2-Na$_x$CoO$_2$ study previously.\cite{Shu2007}  The potential drawback of surface-versus-bulk difference on the transient galvanostatic scan is avoided, in particular, since sensitive magnetic property interpretation can only be addressed when homogeneous Li (Vacancy) contribution for specific single phase is perfectly isolated without ambiguity.  These findings are able to clarify the inconsistency among results obtained through different grain sizes and de-intercalation methods so far.   


\section{\label{sec:level1}Experimental details\protect\\}

A complete series of single crystal Li$_x$CoO$_2$ sample with x $\sim$ 0.33-0.87 were prepared from electrochemical de-intercalation of the pristine single crystal Li$_{0.87}$CoO$_2$.  The pristine single crystal Li$_{0.87}$CoO$_2$ was grown using flux method starting from a mixture of LiCoO$_2$ : Li$_2$O$_2$: LiCl = 1 : 4 : 4, where LiCoO$_2$ was prepared from Li$_2$CO$_3$ and Co$_3$O$_4$ powder with a molar ratio of Li:Co=1:1 and heated at 900 $^\circ$C for 24 hours.  The mixture was sealed in an alumina crucible using alumina cement, heated to 900 $^\circ$C, soaked for 5 hours, slowly cooled to 600 $^\circ$C at a rate of 2 $^\circ$C/hr, and finally furnace cooled to room temperature.  The as-grown single crystal has a typical size of $\sim$5$\times$5$\times$0.1mm$^3$, with Li content of x $\sim$ 0.87 based on inductively coupled plasma-mass spectrometer (ICP) analysis, which is lower than the expected x=1 due to Li vapor loss under the cement sealing condition.  Li content was reduced further through an electrochemical potentiostatic de-intercalation process using various constant applied voltages, where Li$_{0.87}$CoO$_2$ single crystal was used as working electrode, 1M LiClO$_4$ in propylene carbonate as electrolyte, and Platinum as counter and reference electrodes.   All Li contents of Li$_x$CoO$_2$ were examined by ICP method to be within error of $\pm$0.01 and verified using the c-axis vs. x plot.\cite{Amatucci1996}  Lattice parameters were analyzed using Bruker D-8 diffractometer and the magnetic properties were measured using Quantum Design SQUID-VSM.  The single crystal samples of Li$_{0.87}$CoO$_2$ and Li$_{0.5}$CoO$_2$ were further examined using Laue diffraction method with synchrotron X-ray generated from 20 keV at the Taiwan-NSRRC. 

\begin{figure}
\begin{center}
\includegraphics[width=3.5in]{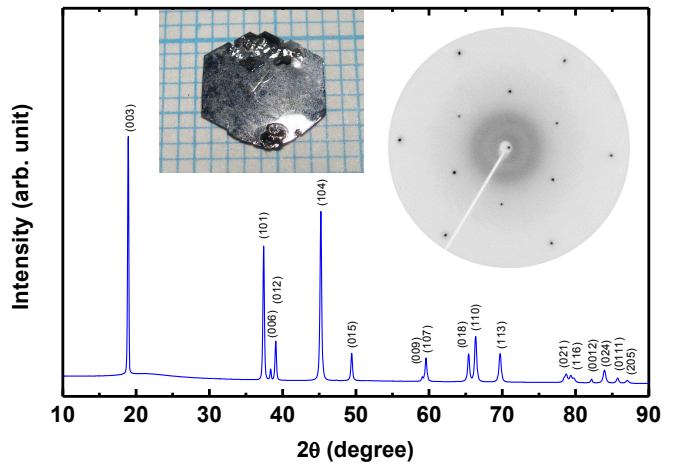}
\end{center}
\caption{\label{fig:fig-XRD}(color online) Synchrotron X-ray powder diffraction pattern for the as-grown single crystal sample Li$_{0.87}$CoO$_2$ which has been ground in powder form.  All diffraction peaks can be indexed correctly using space group R$\bar{3}$m without extra impurity phases identified.  The inset shows the crystal photo and its Laue diffraction pattern of the as-grown single crystal sample.}
\end{figure}

For the purpose of a detailed electronic phase diagram study on Li$_x$CoO$_2$, lithium content and homogeneity has been crucial on its reliability.  Besides the lithium content determined and confirmed through ICP analysis plus lattice parameter comparison as described above, the homogeneity of lithium distribution has been examined carefully using synchrotron X-ray diffraction and spin susceptibility analysis.  The phase purity and crystal structure of the as-grown single crystal sample has been characterized fully using synchrotron X-ray Laue and powder diffraction as shown in Fig.~\ref{fig:fig-XRD}.  The as-grown crystal shows hexagonal morphology which is expected for sample possesses hexagonal CoO$_2$ symmetry within the basal plane.  Synchrotron Laue diffraction shown in the inset of Fig.\ref{fig:fig-XRD} confirmed the R$\bar{3}$m symmetry for the crystal perpendicular to the CoO$_2$ plane with correct index without impurity or superlattice spots.  Lithium homogeneity for each x content has been guaranteed through potentiostatic electrochemical de-intercalation method, where the final induced charge transfer current has been saturated for more than 12 hours at the background level to warrant complete conversion for the whole crystal specimen at the designated OCP.  X-ray diffraction peak width (FWHM) which reflects lithium homogeneity has been confirmed to be maintained within 0.05$\pm$0.01 degree for the whole range of crystal samples studied (not shown).
					
\section{\label{sec:level1}Results and Discussions\protect\\ }

\subsection{\label{sec:level2}Electronic phase diagram\protect\\ }

\begin{figure}
\begin{center}
\includegraphics[width=3.5in]{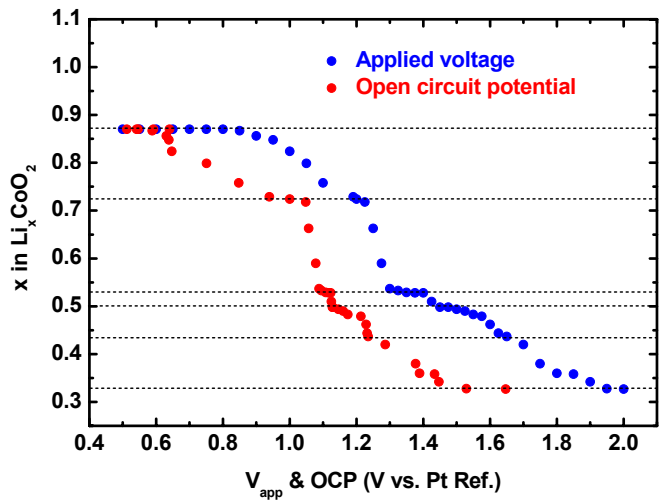}
\end{center}
\caption{\label{fig:fig-xvsV}(color online) Li content x versus applied voltage (V$_{app}$) and open circuit potential (OCP) for Li$_x$CoO$_2$ single crystal electrode.  The Li de-intercalation is controlled using potentiostatic process, i.e., a constant V$_{app}$ (vs. Pt reference) is applied to the sample electrode until the induced current falls to the background level, and OCP is recorded after the charging process has ended for 24 hours.}
\end{figure}

In general, open circuit potential (OCP) represents the chemical potential (relative to the reference electrode potential) of a material when the sample acts as a working electrode in the electrochemical cell.  The applied overpotential above (or below) the current OCP would induce charge transfer between the electrolyte and the working electrode surfaces, and the induced current decays until OCP is tuned to the new level.  The phase diagram mapped by x vs. V as shown in Fig.~\ref{fig:fig-xvsV} reveals several stable phases from the plateau of x vs. V, which closely correlate with the specific Li contents near $\sim$ 0.72, 0.53, 0.50, 0.43, and 0.33.   Although the stacking sequences between alkali metal and CoO$_2$ layers and the space group are different between Na$_x$CoO$_2$ and Li$_x$CoO$_2$, the basic character of the x vs. V plot for Li$_x$CoO$_2$ is very similar to that found for Na$_x$CoO$_2$.\cite{Amatucci1996, Hertz2008, Motohashi2009}  It is not surprising to find phases with x $\sim$ 1/2 and 1/3 when both systems have similar CoO$_2$ networks of 2D triangular lattice.   In addition, in proximity to the exact half filling, there are metastable phases of x $\sim$ 0.53 and 0.43 found for Li$_x$CoO$_2$.  Similar phases have also been identified in Na$_x$CoO$_2$ near x $\sim$ 0.55 and 0.43 as reported earlier.\cite{Shu2007, Shu2008}  

Although P2-Na$_x$CoO$_2$ and O3-Li$_x$CoO$_2$ have different gliding sequences and alkaline metal ion environments as shown in Fig.~\ref{fig:fig-structure}, it is interesting to find that Li$_{0.72}$CoO$_2$ shows particular stability  similar to the fully characterized stable phase of Na$_{0.71}$CoO$_2$.  Na$_{0.71}$CoO$_2$ has been identified by synchrotron Laue X-ray with a proposed simple hexagonal $\sqrt{12}$a$\times$$\sqrt{12}$a$\times$3c superstructure model formed by Na vacancy multivacancy clusters before.\cite{Chou2008, Huang2009, Huang2010}  Na vacancy level that leads to the x=0.71 phase formation is a result of perfect tri- and quadri-vacancy stacking in the space group of P6$_3$/mmc symmetry (or P3$_1$ after Na ordering is considered), i.e., closely related to the necessity of inversion symmetry between the nearest-layer oxygen positions.  While Li$_x$CoO$_2$ has been described in the space group of R$\bar{3}$m with CoO$_2$ layers to glide along the [111] direction of three Li-CoO$_2$ per unit in the hexagonal description as shown in Fig.~\ref{fig:fig-structure}, the Li vacancy level that is responsible for the existence of x $\sim$ 0.72 stable phase is puzzling.  

\begin{figure}
\begin{center}
\includegraphics[width=3.5in]{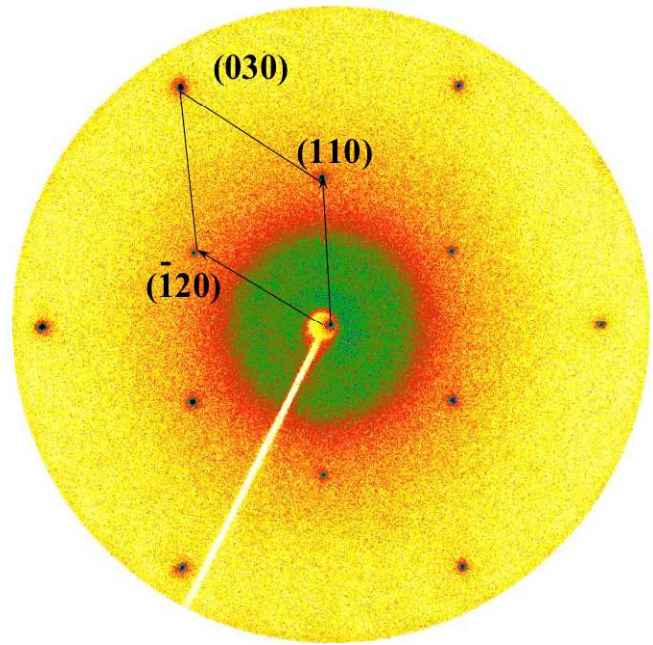}
\end{center}
\caption{\label{fig:fig-Laue}(color online) Laue diffraction pattern for Li$_{0.72}$CoO$_2$ obtained using synchrotron X-ray source of wavelength=0.6199$\AA$ perpendicular to the CoO$_2$ layer at room temperature.}
\end{figure}

To search for the possible Li vacancy ordering, Laue pattern for Li$_{0.72}$CoO$_2$ obtained using synchrotron X-ray is shown in Fig.~\ref{fig:fig-Laue}.  While the transmission Laue is taken perpendicular to the CoO$_2$ layer, a clear hexagonal symmetry can be indexed.  Comparing with the Laue pattern of Na$_{0.71}$CoO$_2$ with well-defined $\sqrt{12}$a hexagonal superlattice,\cite{Chou2008} no superlattice can be identified besides the main low indexing (hk0) planes constructed by the hexagonal Co network of a=2.8156 $\AA$.  Contrary to the Na$_{0.71}$CoO$_2$ superlattice, which is required by the Na trimer formation under P6$_3$/mmc symmetry, Li$_{0.72}$CoO$_2$ has no similar symmetry condition, but, on the other hand, sits close to the phase boundary of the two-phase region between $\sim$0.75-0.94 as reported previously.\cite{Hertz2008, Amatucci1996, Reimers1992}  The phase of relatively shorter c-axis of the two-phase could be related to the partial O2-like gliding as suggested by Carlier $\textit{et al.}$ for the P2- to O2- structural transition.\cite{Carlier2001}  It is possible that the particular stability of Li$_{0.72}$CoO$_2$ may have nothing to do with specific Li vacancy ordering, but rather closely related to the requirement of missing O2-like gliding after the screening effect is weakened by the lower Li level between the CoO$_2$ blocks.  

It is very difficult to obtain single phase Li$_{0.72}$CoO$_2$ using potentiostatic method without going through an extreme fine tuning process on applied voltage, mostly because it sits between the 0.75-0.94 two-phase region (to be discussed in the following sections), and the steep and narrow voltage range for 0.72-0.52 phase formation as shown in Fig.~\ref{fig:fig-xvsV}.  Li-Co chemical disordering has also been ruled out based on the finite-temperature calculation results,\cite{Wolverton1998} although supporting evidence has been found experimentally in the literature.\cite{Menetrier2008} The average diffusion coefficient at room temperature obtained from polycrystalline Li$_x$CoO$_2$ is in the order of D$_{Li}$$\sim$ 10$^{-12}$,\cite{Jang2001} which is nearly four orders lower (D$_{Na}$$\sim$10$^{-8}$ cm$^2$/s) than that of Na$_x$CoO$_2$ estimated from single crystal study,\cite{Shu2008}  except that the minima of D$_{Na}$'s near those specifically stable phases of x $\sim$ 0.71, 0.50, and 0.33 fall to the same average level as D$_{Li}$.  Although the reported D$_{Li}$ values in the literature vary from 10$^{-13}$-10$^{-7}$ and the differences have been attributed to the assumption of geometrical factors used in the calculation,\cite{Jang2001} it is still possible that the specific Li vacancy level of Li$_x$CoO$_2$ does not induce enough chemical potential reduction for potential ionic ordering mechanism, as demonstrated by the missing of Li vacancy ordering for Li$_{0.72}$CoO$_2$.  

Based on OCP studies for Li$_x$CoO$_2$ sample electrode as shown in Fig.\ref{fig:fig-xvsV}, there are two extra phases in proximity to the exact half filling of x = 0.5, i.e., the two distinguishable stable phases revealed by the narrow plateau near $\sim$ 0.53 and 0.43, which have also been observed under galvanostatic scan previously.\cite{Clemencon2007, Jang2001}  These two specific x values are close to those found in Na$_x$CoO$_2$ system near x $\sim$ 0.55 and 0.43 with distinctly different Laue superlattice patterns and magnetic phase transitions.\cite{Shu2007}  The origin of phases with slight deviation from exact x=1/2 remains to be explored, but preliminary electron diffraction studies for  Na$_x$CoO$_2$ with x near 0.5 has been attributed to the metastable phases in proximity to the exact half-filling as a result of alternating vacancy rich/poor 1D Na zigzag chains.\cite{Huang2011a}

\subsection{\label{sec:level2}Magnetic susceptibilities\protect\\ }

\begin{figure}
\begin{center}
\includegraphics[width=3.5in]{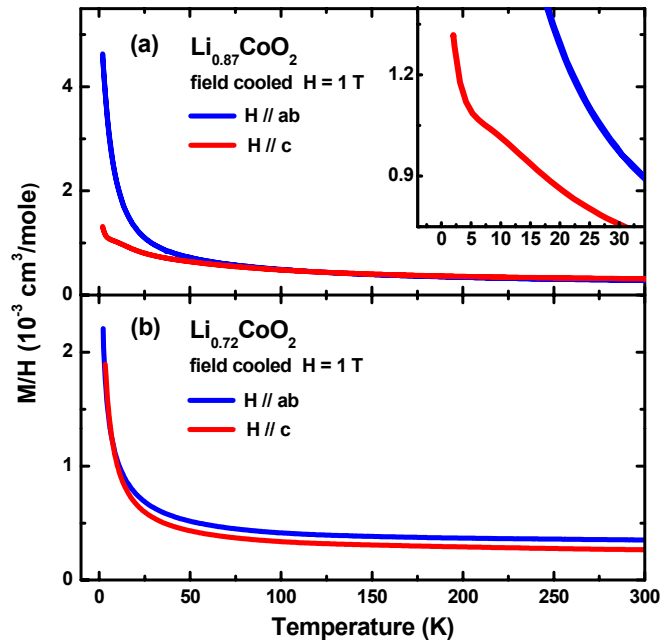}
\end{center}
\caption{\label{fig:fig-chi_highx}(color online) Magnetic susceptibilities for Li$_{0.87}$CoO$_2$ and Li$_{0.72}$CoO$_2$ single crystals.  The inset reveals the anisotropic anomaly near $\sim$10K, which suggests the existence of A-type AF ordering with spins aligned antiferromagnetic along the c-direction.}
\end{figure}

Magnetic susceptibilities for Li$_x$CoO$_2$ (x $\sim$ 0.87, 0.72) single crystals are summarized in Fig.~\ref{fig:fig-chi_highx}.  These results differ significantly from those reported previously based on either powder or single crystal samples prepared with chemical or galvanostatic de-intercalated methods.\cite{Motohashi2009, Miyoshi2010}   We find that the signal size of magnetic anomalies revealed by the sensitive SQUID magnetometry on single crystal samples are very weak, which cannot be reliably observed in polycrystalline samples, especially when inhomogeneity cannot be ruled out in chemical or galvanostatic de-intercalation processes.  Besides, magnetic impurity of Co$_3$O$_4$ may become dominant below 50K under certain annealing conditions and the muon spin resonance tool shows limited sensitivity in this range.\cite{Artemenko2009, Sugiyama2009}   

As shown in Fig.~\ref{fig:fig-chi_highx}(a), the anisotropic magnetic anomaly found near $\sim$10K for the as-grown Li$_{0.87}$CoO$_2$ clearly indicates the existence of possible A-type antiferromagnetic ordering, similar to that found previously in Na$_x$CoO$_2$ for x $\sim$ 0.82-86, where the magnetization is reduced for H$\|$c to signal antiparallel spin arrangement in c-direction.\cite{Shu2007, Shu2009}  The A-type antiferromagnetic (A-AF) ordering for Na$_{0.82}$CoO$_2$ has been confirmed through neutron scattering experiment before,\cite{Bayrakci2004} where the interlayer AF coupling coexists with an in-plane ferromagnetism (FM) of itinerant electron spins.\cite{Chou2008}   The similar finding of A-AF behavior in the as-grown Li$_{0.87}$CoO$_2$ is not surprising, although it has not been observed through magnetic susceptibility measurement before to the best of our knowledge.  

The magnetic susceptibilities for Li$_{0.72}$CoO$_2$ shown in Fig.~\ref{fig:fig-chi_highx}(b) demonstrate similar magnetic behavior to that of Na$_{0.71}$CoO$_2$ also, i.e., no magnetic anomaly has been found between 1.7-300K, which is in great contrast to the finding of 175K phase transition reported previously.\cite{Hertz2008, Miyoshi2010, Sugiyama2009}   Hertz \textit{et al.} have examined the confusing 175K anomaly carefully before and ruled out the possibility of the existence of Co-O impurities.  In addition, the observed aging effect of 175K anomaly strongly implies the occurrence of magnetic moment induced by local microscopic Li inhomogeneity.\cite{Hertz2008}  We suspect that Li inhomogeneity becomes the major problem for samples prepared using chemical or galvanostatic de-intercalation methods.  This can also be argued from the persistent observation of the 175K phase transition with various intensities found in the whole range of x $\sim$ 0.78-0.46, a typical example of possible inclusion of identical impurity phase with strong magnetic signal contribution.\cite{Miyoshi2010, Hertz2008}  On the other hand, our single crystal samples were prepared with well-controlled potentiostatic de-intercalation method to warrant Li homogeneity until there was no difference on the potential reading between surface and the bulk.  Besides, there are two different c-axes detected in the range of $\sim$0.72-1 before single phase formation below $\sim$ 0.72 as reported by various preparation methods and summarized in Fig.~\ref{fig:fig-lattice}.  A first order metal-insulator Mott transition has also been proposed for x $\lesssim$ 0.75 based on density function theory (DFT) calculations, which could be closely related to the two phase coexistence in the same range as a result of gradual de-localization from x $\gtrsim$ 0.95.\cite{Marianetti2004}  Phase purity requirement for the physical property investigation becomes crucial when Li inhomogeneity cannot be controlled easily, especially when electrochemical process can only detect the surface potential.  

Curie-Weiss law fitting for the paramagnetic behavior of the spin susceptibilities for Li$_{0.72}$CoO$_2$ shows that Curie constant is $\sim$ 0.011 cm$^3$-K/mole, which is nearly one tenth of that for Na$_{0.71}$CoO$_2$ reported previously.\cite{Chou2008}  The significantly lower level of localized spins suggests that the doped holes resulting from Li vacancy are mostly mobile.  This observation is consistent with the absence of Li vacancy ordering in the synchrotron Laue photograph, while superlattice ordering of Na vacancy cluster has been proposed to create more localized spins near the vacancy cluster centers and the rest of the doped holes are itinerant.\cite{Chou2008, Balicas2008}  The absence of magnetic ordering for the local moment on a frustrated lattice has been suggested to be a unique spin liquid state for Na$_{0.71}$CoO$_2$. Similar behavior for Li$_{0.72}$CoO$_2$ deserves further investigation also.      

\subsection{\label{sec:level2}Crystal structures\protect\\ }

\begin{figure}
\begin{center}
\includegraphics[width=3.5in]{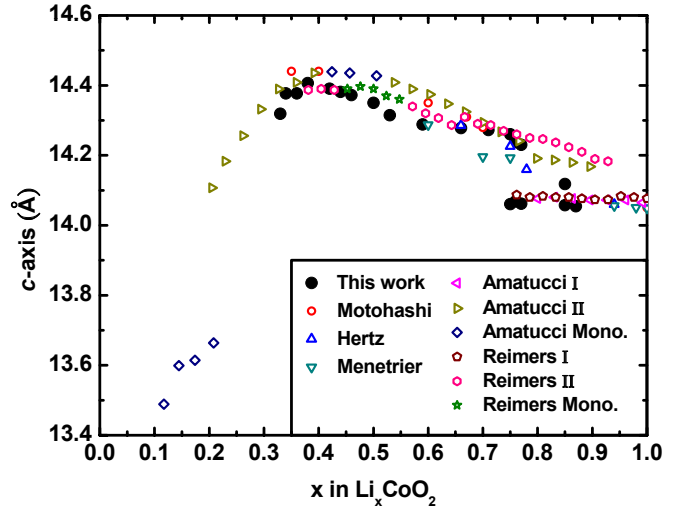}
\end{center}
\caption{\label{fig:fig-lattice}(color online) c-axes for single crystal samples studied in this work (solid circle) are compared with those reported in the literature (empty symbols).\cite{Amatucci1996, Hertz2008, Motohashi2009, Reimers1992, Menetrier1999} Monoclinic  structure distortion has been observed for Li$_x$CoO$_2$ close to x $\sim$ 0.5 and 0.2. }
\end{figure}

The c-axis lattice parameters from five published works plus our current results are summarized in Fig.~\ref{fig:fig-lattice}.\cite{Amatucci1996, Hertz2008, Motohashi2009, Reimers1992, Menetrier1999} We note that the as-grown single crystal Li$_{0.87}$CoO$_2$ falls in the reported two-phase region; however, the as-grown crystal is clearly single phase with smaller c-axis, similar to those for x $\gtrsim$0.95.  The two-phase phenomenon must be coming from the domain-like de-intercalation process before x$\lesssim$0.72 is achieved, which should occur only in the de-intercalation process at room temperature but not through the melt growth condition at high temperatures, as additionally verified by the two-phase observation in our crystal samples of 0.72$\lesssim$x$\lesssim$0.87 obtained by following electrochemical de-intercalation.  The main reason for the persistent two-phase for x$\sim$0.72-1 has been proposed theoretically to be a first order metal-insulator transition as a Mott transition of impurities, where high mobility of Li vacancies allows the $\sim$25$\%$ vacancy metallic phase to grow at the expense of insulating phase of x$\gtrsim$0.95 at room temperature.\cite{Marianetti2004}  It is noted that c-axes values for 0.33$\lesssim$ x $\lesssim$0.72 obtained in this study (Fig.~\ref{fig:fig-lattice}) are consistently lower than those  widely distributed values reported previously.\cite{Amatucci1996, Reimers1992}  Li inhomogeneity in powder samples is a reasonable assumption to explain the difference, i.e., it is highly likely that the Li contents were over-estimated before as a result of surface-to-bulk potential difference during de-intercalation process for the powder sample, i.e., possibly with mixture of low x phases which are closer to x$\lesssim$0.5 at the grain boundaries, especially when galvanostatic de-intercalation route has been applied.

\subsection{\label{sec:level2}Monoclinic Li$_{0.5}$CoO$_2$\protect\\ }

\begin{figure}
\begin{center}
\includegraphics[width=3in]{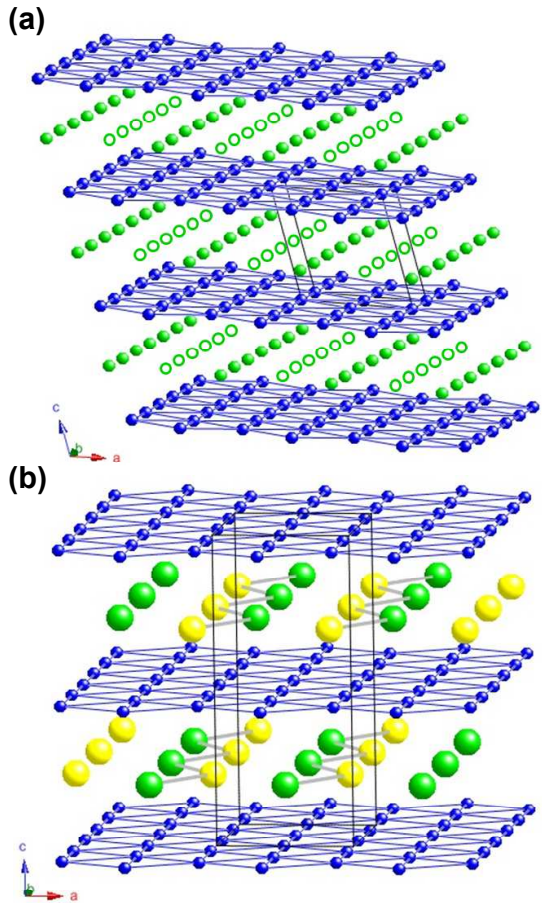}
\end{center}
\caption{\label{fig:fig-Li05structure}(color online) Crystal structures of (a) Li$_{0.5}$CoO$_2$ described in P2/m,\cite{Shao-Horn2003} where solid and empty circles in green represent Li and Li-vacancy respectively, and (b) Na$_{0.5}$CoO$_2$ with Na sublattice described in space group of Pnmm\cite{Huang2004}, where zigzag chains formed by Na1 (green) and Na2 (yellow) solids have been defined in the original P6$_3$/mmc symmetry.}
\end{figure}

Monoclinic distortion has been observed previously in Li$_x$CoO$_2$ close to x$\sim$0.5 and 0.2 at room temperatures.\cite{Amatucci1996, Reimers1992}  The crystal structure of Li$_{0.5}$CoO$_2$ has been identified as monoclinic lattice with space group $P2/m$.\cite{Takahashi2007, Shao-Horn2003}  The gliding blocks of CoO$_2$ and Li for Li$_x$CoO$_2$ with R$\bar{3}$m symmetry of three Li-CoO$_2$ layers per unit must favor the monoclinic distortion when Li content is reduced to half.  In addition, the alternating linear Li-vacancy chains for Li$_{0.5}$CoO$_2$ is quite different from the zigzag Na chain in Na$_{0.5}$CoO$_2$.\cite{Huang2004}  Li in O3-Li$_x$CoO$_2$ has only one site, which sits directly beneath the Co column, similar to the Na1 site in P2-Na$_x$CoO$_2$.\cite{Chou2008}  But when cation vacancy is generated through de-intercalation, Na ion can move from the originally energetically favorable Na2 site to the now  energetically more favorable  Na1 site to form Na-trimer or multi-vacancy cluster,\cite{Roger2007} which is not a possible mechanism under Li$_x$CoO$_2$ R$\overline{3}$m symmetry, and an alternative Monoclinic distortion mechanism is chosen.  In fact, mazed domain structure of monoclinic Li$_{0.5}$CoO$_2$ has been revealed through electron diffraction before, where equivalent domains of in-plane Li-vacancy ordering and other nm size domains of variants exist.\cite{Shao-Horn2003}

\begin{figure}
\begin{center}
\includegraphics[width=3.5in]{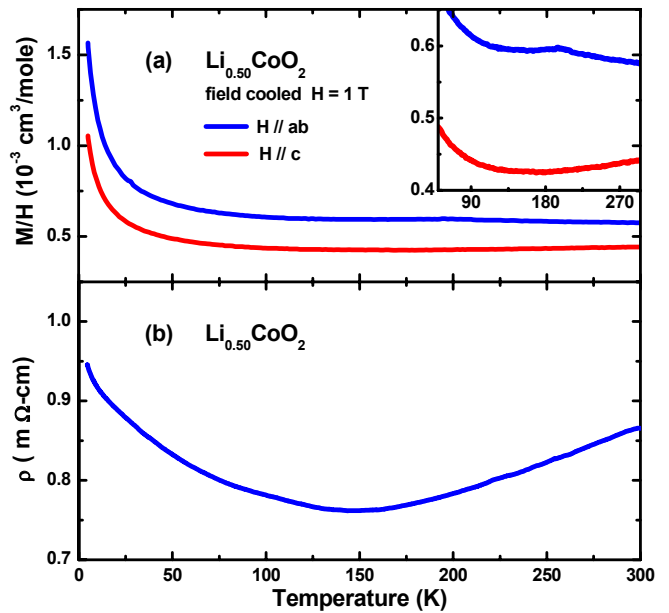}
\end{center}
\caption{\label{fig:fig-x05}(color online) (a) Magnetic susceptibility for Li$_{0.5}$CoO$_2$ single crystal measured using 1 Tesla for H$\|$ab and H$\|$c, and the transition region near $\sim$200K enlarged in the inset.  (b) The in-plane resistivity for x=0.5.}
\end{figure}

The magnetic susceptibility and resistivity measurement results for Li$_{0.5}$CoO$_2$ are shown in Fig.~\ref{fig:fig-x05}.  There is a small dip of susceptibility found near $\sim$ 200K for H$\|$ab only as revealed in the inset of Fig.~\ref{fig:fig-x05}(a), but no corresponding anomaly found for H$\|$c, which is similar to the behavior of antiferromagnetic transition near $\sim$ 88K found in Na$_{0.5}$CoO$_2$.  This AF spin ordering for Na$_{0.5}$CoO$_2$ has been proposed to be coming from the alternating rows of AF ordered and nonordered Co ions within the ab-plane.\cite{Gasparovic2006}  Columns of alternating Li-vacancy for Li$_{0.5}$CoO$_2$ is shown in Fig.~\ref{fig:fig-Li05structure}(a).\cite{Shao-Horn2003, Takahashi2007}  The anisotropic anomaly near $\sim$ 200K for Li$_{0.5}$CoO$_2$ could be coming from a similar AF ordering, and spin polarized neutron scattering should be performed to check this possibility.  For Na$_{0.5}$CoO$_2$ shown in Fig.~\ref{fig:fig-Li05structure}(b), the intricate Co columns sandwiched between upper and lower zigzag Na chains must form the ordered Co with AF spins, while the in-between Co columns are spinless as supported by the neutron scattering study results.\cite{Gasparovic2006}  Similarly, Li$_{0.5}$CoO$_2$ also possesses alternating Li-vacancy columns and the nearby Co columns must generate similar ordered AF ordered spins and non-ordered spinless columns in the hexagonal Co-plane.  

Although there is a significant metal-to-insulator transition follows at $\sim$51K for Na$_{0.5}$CoO$_2$, there is only a very weak increase of resistivity found below $\sim$150K for Li$_{0.5}$CoO$_2$.  The metal-insulator transition below $\sim$ 51K for Na$_{0.5}$CoO$_2$ has been attributed to the charge localization as a result of delicate balance between local Coulomb repulsion and kinetic energy in the CoO$_2$ plane when ideal interpenetrating filled (spinless) and half-filled (with spin) orthorhombic sub-lattices are constructed.\cite{Choy2007}  On the other hand, there is no clear metal-to-insulator transition observed down to 4.2K for Li$_{0.5}$CoO$_2$, except a resistivity minimum occurs near $\sim$150K as shown in Fig.~\ref{fig:fig-x05}(b).  Unlike the maintained Na zigzag chain ordering along the c-direction for Na$_{0.5}$CoO$_2$, the monoclinic distortion for Li$_{0.5}$CoO$_2$ along c-axis must be responsible for the missing charge ordering when 3D ordering along c-direction is broken.

\subsection{\label{sec:level2} De-intercalation and Li homogeneity\protect\\ }

\begin{figure}
\begin{center}
\includegraphics[width=3.5in]{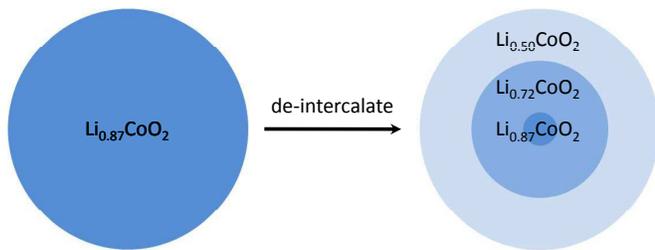}
\end{center}
\caption{\label{fig:fig-radial}(color online) Radial model of inhomogeneous galvanostatic or chemical de-intercalation for Li$_x$CoO$_2$. }
\end{figure}

\begin{table*}
\begin{center}
\caption{\label{tab:tableI} Summary of Li$_x$CoO$_2$ samples prepared with different de-intercalation routes}
\begin{tabular}{cccc}   
\hline
Reference & deintercalation method  & powder/crystal & 175K anomaly\\
\hline\hline
Imanishi \textit{et al.} 1999\cite{Imanishi1999} & chemical & powder & x$\sim$0.76-0.61\\
Menetrier \textit{et al.} 1999\cite{Menetrier1999} & EC galvanostatic & powder & N.A.\\
Clemencon \textit{et al.} 2007\cite{Clemencon2007} & EC galvanostatic & powder & N.A.\\
Mukai \textit{et al.} 2007\cite{Mukai2007} & EC galvanostatic & powder & x$\sim$0.75-0.50 \\
Hertz \textit{et al.} 2008\cite{Hertz2008} & chemical & powder & x$\sim$0.78-0.51 \\
Kawasaki \textit{et al.} 2009\cite{Kawasaki2009} & EC galvanostatic & powder & N.A.\\
Sugiyama \textit{et al.} 2009\cite{Sugiyama2009} & EC galvanostatic & powder & x$\sim$0.73-0.53\\
Motohashi \textit{et al.} 2009\cite{Motohashi2009} & EC galvanostatic & powder & x$\sim$0.70-0.50\\
Mohanty \textit{et al.} 2009\cite{Mohanty2009} & chemical & powder & no\\
Miyoshi \textit{et al.} 2010\cite{Miyoshi2010} & ion exchange+chemical & single crystal & x$\sim$0.71-0.46\\
Ishida \textit{et al.} 2010\cite{Ishida2010} & ion exchange+chemical & thin film & x$\sim$0.66\\
\hline
This work  & EC potentiostatic & single crystal & no\\
\hline\hline
\end{tabular}\end{center}
\end{table*}

It has been puzzling to find that there is one persistent phase that can be identified through several experimental techniques near $\sim$150-175K within a wide range of x $\sim$ 046-0.78 as summarized in Table~\ref{tab:tableI}.  In addition, the observed 175K phase anomaly shows a significant thermal hysteresis in both resistivity and magnetic susceptibility measurements.\cite{Miyoshi2010, Motohashi2009}  The persistent existence of one phase signature in a wide range of lithium content suggests that such phase has not been isolated completely from the co-existing phase mixture.  The observation of thermal hysteresis also strongly suggests that such hard-to-extract phase could be closely related to a phase existence within extremely narrow phase space, i.e., slight deviation due to inhomogeneity could create phase separation of small miscibility gap.  In fact, Hertz \textit{et al.} have proposed that the detected 175K anomaly could be associated with a phase that can only be stabilized at the surface or grain boundaries, as suggested by its aging character and mostly found in powder samples.\cite{Hertz2008}  Examine the electronic phase diagram mapped by the potentiostatic de-intercalation method as shown in Fig.~\ref{fig:fig-xvsV}, we find that indeed there exists a sharp x vs V slope within $\sim$20 mV for sample preparation of x$\sim$0.52-0.72, i.e., small deviation on applied potential would produce two-phase coexistence unless post annealing at 120$^\circ$C is applied.   We failed to isolate the expected phase that shows similar 175K anomaly after repeated careful tuning of V$_{app}$ near $\sim$125 mV following the electronic phase diagram shown in Fig.~\ref{fig:fig-xvsV},.  Although these negative results cannot exclude the existence of a phase that shows $\sim$175K anomaly as revealed by several experimental techniques, which strongly supports that this peculiar phase must exist within a extremely narrow phase space to be isolated cleanly or associate with surface and grain boundaries closely.     


Most early studies of Li$_x$CoO$_2$ phase diagram used galvanostatic scans on polycrystalline samples in  electrochemical battery cells, and the stable phases were isolated using the characteristic surface potential and X-ray diffraction.  The battery construction using fine polycrystalline sample is efficient for revealing the occurrence of stable phases through surface potential with total charge estimated using Faraday's law.  Alternatively, chemical de-intercalation using oxidant under various concentrations can also serve the same purpose, as summarized in Table~\ref{tab:tableI}.  However, these processes are of approximation for a quasi-equilibrium process at the limited time scale, i.e., unless the charging rate of galvanostatic current density is lower than the Li diffusion rate, and the charging process can be stopped immediately after reaching the targeted levels.  Strictly speaking, galvanostatic de-intercalation method cannot guarantee single phase formation with accurate charge integration, especially when multiple stable phases continue to build up on the outer layers of the particle, yet only the surface potential is recorded.  Radial and Mosaic models have been proposed to describe the intercalation problem for the battery electrode material Li$_x$FePO$_4$ with only two stable phases of x = 1 and 0,\cite{Andersson2001}  but there are more stable phases identified in Li$_x$CoO$_2$ as proposed in Fig.~\ref{fig:fig-radial} following a Radial model description.  It is demonstrated that the delithiation process cannot be a complete process in the galvanostatic scan, i.e., there is always remanent stable phases in the inner core area that cannot be reached in time galvanostatically when the charging current is terminated by the achieved surface potential before diffusion process is completed.


\section{\label{sec:level1}Conclusions\protect\\ }

In summary, by using a series of potentiostatically de-intercalated single crystal Li$_x$CoO$_2$, we find great similarity between electronic phase diagrams of Na$_x$CoO$_2$ and Li$_x$CoO$_2$.  Contrary to most reports based on powder sample prepared with galvanostatic or chemical de-intercalation methods, current work revealed an A-type antiferromagnetic signature for x $\sim$ 0.87 and the non-existence of magnetic ordering down to 1.7K for x $\sim$ 0.72.  Although Li$_{0.72}$CoO$_2$ shows similar stoichiometry and particular stability compared to Na$_{0.71}$CoO$_2$ with well-defined Na vacancy superlattice ordering, the former does not possess any Li vacancy ordering.  Near half filling of Li, three phases of x $\sim$ 0.53, 0.50, and 0.43 have been identified and isolated to show distinct physical properties.  Comparing with Na$_{0.5}$CoO$_2$, no metal-insulator transition has been found for Li$_{0.5}$CoO$_2$, although similar  antiferromagnetic ordering is implied below $\sim$ 200K.  Based on a Radial model interpretation, current  potentiostatic study results based on single crystal samples support that the widely observed within x $\sim$ 0.75-0.50 but hard-to-extract phase with a magnetic anomaly near $\sim$ 175K could be related to a minor phase sitting preferentially on the surface or grain boundaries.  



\section*{Acknowledgment}
FCC acknowledges the support from National Science Council of Taiwan under project number NSC-99-2119- M-002-011-MY2. \\
\appendix


\end{document}